\documentstyle{l-aa}

\def\Cb{\overline{C}}
\def\Db{\overline{D}}
\def\M0b{\overline{M}_0}
\def\disp{\displaystyle}

\def\d{{\rm d}}

\def\mD{{\cal D}}

\def\vx{{\vec x}}
\def\vk{{\vec k}}
\def\vl{{\vec l}}
\def\ii{{\rm i}}
\def\em{\sl}
\def\gam{{\vec{\gamma}}}

\def\mg{\big <}
\def\md{\big >}

\def\ort{\perp}
\def\gamo{\vec{\gamma}^{\rm obs.}}
\def\gamt{\vec{\gamma}^{\rm prim.}}
\def\dgam{\vec{\delta}\vec{\gamma}}
\def\dTo{\delta_T^{\rm obs.}}
\def\dTt{\delta_T^{\rm prim.}}
\def\chicmb{\chi_{\rm CMB}}
\def\be{\begin{equation}}
\def\ee{\end{equation}}
\def\ba{\begin{eqnarray}}
\def\ea{\end{eqnarray}}

\input psfig.sty

\begin{document}

   \thesaurus{12 (12.03.1; 12.04.1; 12.07.1; 12.12.1)} 

 \title{Weak Lensing Detection in CMB Maps}

 \author{F. Bernardeau}

 \offprints{F. Bernardeau; fbernardeau@cea.fr}

 \institute{Service de Physique Th\'eorique, 
C.E. de Saclay, F-91191 Gif-sur-Yvette cedex, France\\}

\maketitle

\markboth{Weak Lensing in CMB}{F. Bernardeau}

\begin{abstract}

The weak lensing effects are known to change only weakly
the shape of the power spectrum of the Cosmic Microwave Background (CMB) 
temperature fluctuations. 
I show here that they nonetheless induce specific non-Gaussian effects
that can be detectable with the {\em four-point correlation function}
of the CMB anisotropies. The magnitude and geometrical dependences 
of this correlation function are investigated in detail.
It is thus found to scale as the square of the derivative
of the two-point correlation function and as the angular correlation
function of the gravitational displacement field.
It also contains specific dependences on the shape of the
quadrangle formed by the four directions.

When averaged at a given scale, the
four-point function, that identifies with the
connected part of the fourth moment 
of the probability distribution function of the local
filtered temperature, scales as the square of
logarithmic slope of its second moment,
and as the variance of the gravitational magnification
at the same angular scale. 

All these effects have been computed for specific
cosmological models. It is worth noting that, as the amplitude 
of the gravitational lens effects has
a specific dependence on the cosmological parameters,
the detection of the four-point correlation function
could provide precious complementary constraints to those
brought by the temperature power spectrum.


\end{abstract}

\section{Introduction}

A robust prediction of inflationary scenarios is that the temperature
fluctuations of the Cosmic Microwave Background (CMB)
are expected to obey Gaussian
statistics. Actually this prediction has been challenged
recently by several authors 
(Falk, Rangarajan \& Srednicki 1993, Munshi, Souradeep \& Starobinsky 1995) 
who calculated the skewness induced by 
nonlinear couplings in the primary\footnote{``primary'' 
means the anisotropies induced  on the last scattering surface by either 
potential fluctuations, Doppler effects or photon density fluctuations, 
not taking into account the secondary effects nor the foregrounds.} 
stage of the temperature fluctuation generation.  The skewness induced 
at this level  has been found, however, to be entirely negligible compared 
to the cosmic variance, and thus not accessible to detections.
Therefore, the primary temperature maps 
are entirely defined, in a statistical sense, by the power spectrum
of the temperature fluctuations or equivalently by the
shape of the two-point correlation function. 
A number of other statistical indicators are thus set up 
by this a priori hypothesis. In particular 
Bond \& Efstathiou
(1987) have investigated expected properties of such temperature
maps, as the number density of temperature peaks, their correlation
functions... Moreover,
the exploration of the CMB physics has been boosted recently
after it has been realized that it would be
possible to determine all the cosmological parameters with a
remarkable precision from
an accurate, and accessible, measurement of the temperature power spectrum 
(Jungman et al. 1996).
In particular, the effects of the secondary sources of temperature
fluctuations (Sunyaev-Zel'dovich effects,
nonlinear Doppler effects, lenses..) and foregrounds (point sources,
galactic dust) on the power spectrum
have been investigated in more details (see Cobras/Samba report,
1996, for a general discussion on these problems). These calculations have
shown that most, if not all, of these effects have a relatively small
impact on it. In these calculations, however, the impact 
of secondary effects on the Gaussian nature of the temperature
field has not been considered.
Particularly interesting are the the higher order correlation functions
that are identically zero for Gaussian  fields, and are thus direct
indicators of any, even small, non-Gaussian features.

In this paper the calculations will be focused on the effects of weak-lensing
on CMB maps. They, indeed, constitute a particularly attractive
mechanism because it comes from a {\sl coupling} 
between the primary temperature fluctuation field 
and the mass concentration on the line of sight acting as deflectors.
Their impact on the power spectrum has been investigated primarily
by Blanchard \& Schneider, (1987) who found the effect to be
negligible. More recent works (Kashlinsky 1988,
Cole \& Efstathiou 1989, Sasaki 1989, Tomita \& Watanabe 1989, 
Linder 1990, Cay\'on, Mart\'\i nez-Gonz\'alez \& Sanz 1993a, b,
Fukushige, Makino \& Ebisuzaki 1994, Seljak 1996)
eventually confirmed this conclusion, although the point was debated
for a while. 
In particular Seljak (1996) made a detailed calculations
of these effects for realistic models of CMB anisotropies
and using a semi-analytic
calculation based on a power spectrum approach that includes
nonlinear corrections. 
In this text I will follow a rather similar approach to investigate
the apparition of non-Gaussian features caused by weak lensing effects.

In section 2, I present the basis of the physical
mechanisms describing the CMB map deformations
induced by weak lensing effects. In Section 3, I give the 
explicit expression of the first non-vanishing correlation function,
the four-point, in different remarkable geometries.
In Section 4, quantitative predictions are given for two different
cosmological models. The dependence of the results on the
cosmological parameters, and the practical interests that such
a measurement could have, are discussed in the last section.

\section{Weak Lensing Effects on CMB Maps}

\subsection{The Basis of the Physical Mechanism}

The effect of a gravitational lens is to induce 
a displacement of the light path, thus moving the apparent
position of a sky patch on the last scattering surface by 
a given angle. The temperature of this patch is not
affected itself, i.e. lenses do not created
new structures, and a perfectly isotropic sky would remain so.
The patch of the sky observed at the position $\gamo$ 
is thus actually coming from the position $\gamt$ on the 
``primordial sky'', and the displacement, $\dgam$, is induced 
by the mass concentration on the line of sight. More precisely
$\dgam$ is given by the transverse derivative of the projected
potential $\phi$ of the mass fluctuations,
\ba
\dgam\equiv&\gamt(\gam)-\gamo(\gam)\nonumber\\
=&\disp{2\int_0^{\chicmb}\d\chi\,
\disp{\mD_0(\chicmb,\chi)\over \mD_0(\chicmb)}
\nabla_{\ort}\phi(\chi)},
\ea
where $\mD_0$ is the angular distance, 
$\chi$ is the distance of the lenses along the line of sight
and $\chicmb$ is the distance of the last scattering surface
(see Kaiser 1992, 
Seljak 1996, Bernardeau, van Waerbeke \& Mellier 1996 for more details
on this equation).
It is interesting to rewrite this equation in terms
of the Fourier transform of the {\sl mass density fluctuation} field.
The Fourier transforms $\delta(\vk)$ are defined by,
\ba
\delta(\gam,\chi)=\int{\d^3 \vk\over (2\,\pi)^{3/2}}
\ D_+(\chi)\ \delta(\vk)\times\nonumber\\
\exp\left[\ii \mD_0\,\vk_{\ort}\cdot\gam+\ii k_r\,\chi\right],
\label{dk}
\ea
where the linear growth factor $D_+(\chi)$ and the Fourier transforms
are normalized to the present time. Then the potential reads,
\ba
&\phi(\gam,\chi)=\disp{3\over 2}\Omega_0\ \disp{D_+(\chi)\over a(\chi)}\,
\int\disp{\d^3\vk\over (2\,\pi)^{3/2}}\disp{\delta(\vk)\over k^2}\times\nonumber\\
&\exp\left[\ii \mD_0\,\vk_{\ort}\cdot\gam+\ii k_r\,\chi\right]
\ea
which implies that the displacement can be written,
\ba
\dgam&=\disp{
\int_0^{\chicmb}\d\chi\ w(\chi)\int\disp{\d^3\vk\over (2\,\pi)^{3/2}}}
\times\label{dgam}\\
&\disp{\ii\,\vk_{\ort}\over k^2\,\mD_0(\chi)}\ \delta(\vk)\,
\exp\left[
\ii \mD_0(\chi)\,\vk_{\ort}\cdot\gam+\ii k_r\,\chi\right],\nonumber
\ea
with
\be
w(\chi)=3\Omega_0\,
\disp{\mD_0(\chicmb-\chi)\,\mD_0(\chi)\over\mD_0(\chicmb)}
\disp{{D_+(\chi)\over a(\chi)}}.\label{eff}
\ee
The function $w(\chi)$ gives the efficiency function of lenses for
sources located on the last scattering surface.
It will be investigated in more details in the last section.

Note that in the following I will amply
use the small angle approximation. It implies in particular
that a given patch of the sky can be decomposed in flat
waves and also that, in moment calculations, the
component of $\vk$ along the line of sight can be neglected
compared to the norm of $\vk_{\ort}$.

\subsection{The Effects on CMB Maps}

Compared to detections on background galaxies, the investigation of lens
effects on the last scattering surface is very attractive, because this
surface is 
at a well defined redshift, and has a negligible width. The analysis
of the lens effects requires however more sophisticated tools since
the induced shear cannot be directly measured. 
The primordial temperature patches on the CMB sky are indeed
known only statistically and have a large angular correlation length.
In which way, then, can the lens effects be revealed? 
Actually lensed CMB maps can be seen
as collections of temperature patches of different sizes and
shapes, which 
or only a fraction of which are displaced or deformed. 
Although this is slightly arbitrary, two effects can be
distinguished in the way sizes and shapes of patches are affected,
\begin{itemize}
\item{} the {\em shear} effect that deforms, stretches out temperature 
patches in the shear direction,
\item{} the {\em magnification} effect that globally enlarges or
shrinks those patches.
\end{itemize}
The local deformations of the temperature patches are however a priori 
difficult to disentangle from the actual primordial
intrinsic temperature fluctuations\footnote{In particular there are 
no known working method to construct weak-lensing maps from 
CMB temperature maps.}.
What will make then the effects detectable is the fact that
{\em close} patches will be deformed in a {\em similar}
way (when they are seen through a unique lens), and the excess
of these close rare features cannot be accounted from a Gaussian field.
It is thus possible to quantify their presence 
by statistical indicators. The power spectrum is of
course not adapted to take into account the
apparition of such non-Gaussian features. 
For that matter the high-order correlation functions,
that are all identically zero for pure Gaussian fields, are
extremely precious. Indeed these higher-order correlation
functions contain informations about shapes, and their derivations
can be pursued completely with Perturbation
Theory techniques. In the following I focus my analysis
on the first non vanishing correlation function, 
the four-point one.

\section{The Effects of Weak Lensing on Temperature Correlation Functions}

\subsection{Statistical Properties}

From the first equation it is easy to see that the temperature, $\dTo$,
observed in the direction $\gam$ is in fact the unaffected
temperature $\dTt$ coming from a slightly changed direction,
\be
\dTo(\gamo)=\dTt(\gamt)=\dTt(\gamo+\dgam).\label{dto}
\ee
In the following I assume that the displacement is small
compared to the angular scale at which the observations are made.
This is a fair assumption since the displacement 
is at most of $1'$ (for cores of clusters) and that the
angular resolution of the future satellite missions does not go below $10'$.
As a result it is always possible to expand the relation (\ref{dto})
with respect to the displacement,
\ba
\dTo(\gam)&=\dTt(\gam)+
\disp{\partial \over \partial \gam_i}\dTt(\gam)\dgam_i+\nonumber\\
&\disp{\partial^2 \over 
\partial \gam_i \partial \gam_j}\dTt(\gam)\dgam_i\dgam_j+\dots\label{exp}
\ea
where the Einstein index summation prescription is used.

It is important to have in mind that both quantities $\dTt(\gam)$
and $\dgam$ are {\sl independent} Gaussian fields. The
primordial temperature field is Gaussian for inflationary scenario,
and independent of the lens potential field because the last scattering
surface is far from the intervening lenses (that will be found
to be at redshift $\la 5$). 
The lens potential field will be assumed to be in the linear regime, although
this is not a crucial hypothesis\footnote{This assumption allows the use 
of the linear power spectrum, but one could have included nonlinear 
corrections to it, since the lens density field is by no means required to 
be Gaussian.}.

In the following the correlation functions or moments will be calculated in 
the small angle approximation, for which the plane approximation
for the last scattering surface can be made. Thus one can write,
\be
\dTt(\gam)=\int\disp{\d^2\vl\over (2\pi)}\,a_l\,\exp(\ii\vl\!\cdot\!\gam),
\ee
where the $a_l$ coefficients obey Gaussian statistical rules.
In particular, 
\be
\mg a_l\,a_{l'}\md=\delta_{\rm Dirac}(\vl+\vl')\ C_l,\label{powl}
\ee
where the $C_l$ are the ``famous'' $C_l$ describing the angular power
spectrum.

On the other hand the random variables $\delta(\vk)$ 
obey the statistics,
\be
\mg\delta(\vk)\,\delta(\vk')\md=\delta_{\rm Dirac}(\vk+\vk')\,P(k),\label{powk}
\ee
where $P(k)$ is normalized to the present day.
In order to produce a consistent set of power spectra it is important to 
have a consistent normalization for $C_l$ and $P(k)$.
This can be obtained from the small $l$ behavior of $C_l$
corresponding to the small $k$ behavior of $P(k)$. In the following I
will assume an Harrison-Zel'dovich initial spectrum, so that, 
\be
P(k)=A\ k\ \ \ {\rm at\ small}\ k. 
\ee
The coefficient $A$ can be related to the small $l$ behavior
of $l$ (see Hu 1995 for instance),
\be
l(l+1)\,C_l=\disp{A\over 4\pi}\,\Omega_0^2\,
\disp{\left({a(\chicmb)\over D_+(\chicmb)}\right)^2}\ \ \ {\rm at\ small}\ l. 
\label{norm}
\ee

\subsection{The Two-Point Correlation Function}

The dominant corrective term for
the angular two-point correlation function can be calculated
from the expansion (\ref{exp}),
\ba
&\mg\dTo(\gam_1)\dTo(\gam2)\md=
\mg\dTt(\gam_1)\dTt(\gam2)\md+\label{corr2}\\
&\mg
\disp{\partial \over \partial \gam_i}\dTt(\gam_1)\dgam_i(\gam_1)\ 
\disp{\partial \over \partial \gam_j}\dTt(\gam_2)\dgam_j(\gam_2)
\md+\nonumber\\
&\mg\dTt(\gam_1)
\disp{\partial^2 \over 
\partial \gam_i \partial \gam_j}\dTt(\gam_2)\dgam_i(\gam_2)\dgam_j(\gam_2)\md+
\dots\nonumber
\ea
The corrective terms have been written up to the quadratic
term in the large-scale structure density field.
The previous expression can be written in Fourier space,
\ba
&\mg\dTo(\gam_1)\dTo(\gam2)\md=\disp{\int
{\d^2\vl\over (2\pi)^2}\ C_l\ \exp\left[\ii\vl\!\cdot\!\gam_{12}\right]+}\\
&\disp{\int{\d^2\vl\over (2\pi)^2}\ C_l\ \exp[\ii\vl\!\cdot\!\gam_{12}]
\int\d\chi_1\d\chi_2\,w(\chi_1)\,w(\chi_2)\times}\nonumber\\
&\disp{\int{\d^3\vk\over(2\pi)^3}}\ 
P(k)\ \disp{\left({\vl\cdot\vk\over k^2\,\mD_0(\chi_1)\,\mD_0(\chi_2)}
\right)^2}\exp[\ii(\chi_1-\chi_2)k_r]
\times\nonumber\\
&\disp{\big(\exp[\ii\mD_0(\chi_1)\vk_{\ort}\!\cdot\!\gam_1-
\ii\mD_0(\chi_2)\vk_{\ort}\!\cdot\!\gam_2]-2\big)}\nonumber
\ea
where $\gam_{12}$ is the angular distance between $\gam_1$ and
$\gam_2$, $\gam_{12}=\vert\gam_1-\gam_2\vert$. This formula has been 
obtained using eqs. (\ref{powl},\ref{powk}) defining the two power spectra.
To complete  the calculations one can use the 
small-angle approximation, well verified below $1$ degree scale, 
which implies that
\be
k\approx k_{\ort}\ \ {\rm and}\ \ P(k)\approx P(k_{\ort}).
\ee
Then the integral over $k_r$ leads to a Dirac function in
$\chi_1-\chi_2$. We eventually have, 
\ba
&\mg\dTo(\gam_1)\dTo(\gam2)\md=C(\gam_{12})+\\
&\disp{\int{l\,\d l\over 2\pi}}\,\int\d\chi\ w^2(\chi)\,
\int{k\,\d k\over 2\pi}\,P(k)\,\disp{l^2\over k^2\,\mD_0^2(\chi)}\times\nonumber\\
&\left[\left(J_0(k\mD_0\,\gam_{12})-1\right)J_0(l\,\gam_{12})+
J_2(k\mD_0\,\gam_{12})\,J_2(l\,\gam_{12})\right],\nonumber
\ea
with
\be
C(\gam)=\disp{\int{l\,\d l\over 2\pi}\ C_l\ J_0(l\,\gam_{12})}.\label{cgam}
\ee
This result does 
not coincide apparently with the one of Seljak (1996, equation A.6),
but simply because the exponential was not expanded
in his expression. That this expansion
can be done is amply justified by the fact that the displacements
are small compared to the angular resolution scale.
Then one recovers exactly the same expression.
Note that the autocorrelation function of the displacement 
field is automatically introduced by the third term in (\ref{corr2}).

\subsection{Higher-Order Correlation Function}

It is quite easy to see that the weak-lensing effects do not
introduce a three-point correlation function. It is indeed
impossible to build a term of non-zero ensemble average 
involving three $a_l$ factors.

The first non trivial high order correlation function is thus the
four-point correlation function.
At this stage it is important to have in mind that
the observable quantity is the {\sl connected} 
part, 
$\mg\dTo(\gam_1)\,\dTo(\gam_2)\,\dTo(\gam_3)\,\dTo(\gam_4)\md_c$,
of the ensemble average,
$\mg\dTo(\gam_1)\,\dTo(\gam_2)\,\dTo(\gam_3)\,\dTo(\gam_4)\md$, that
is the part which is obtained when the products of two point correlation
functions that can be built have been subtracted out,
\ba
&\disp{
\mg\dTo(\gam_1)\,\dTo(\gam_2)\,\dTo(\gam_3)\,\dTo(\gam_4)\md_c\equiv}\\
&\disp{
\mg\dTo(\gam_1)\,\dTo(\gam_2)\,\dTo(\gam_3)\,\dTo(\gam_4)\md}-\nonumber\\
&\disp{
\mg\dTo(\gam_1)\,\dTo(\gam_2)\md\,\mg\dTo(\gam_3)\,\dTo(\gam_4)\md}
-\nonumber\\
&{\rm perm.}\ \ (2\ {\rm other\ terms}).\nonumber
\ea
The connected part is obviously zero for the primordial
field: it is a direct consequence of its Gaussian nature. The
dominant term, in terms of weak lensing effects, is thus given by,
\ba
&\disp{
\mg\dTo(\gam_1)\,\dTo(\gam_2)\,\dTo(\gam_3)\,\dTo(\gam_4)\md_c\equiv}
\label{expc4}\\
&\mg\dTo(\gam_1)\ 
\disp{\partial \over \partial \gam_i}\dTt(\gam_2)\md
\mg\dgam_i(\gam_2)\ \dgam_j(\gam_3)\md\times
\nonumber\\
&\mg\disp{\partial \over \partial \gam_j}\dTt(\gam_3)\ 
\dTo(\gam_4)\md+{\rm perm.}\ (11\ {\rm other}\ {\rm terms}).\nonumber
\ea
Roughly speaking it means that the four-point correlation
function, in units of the square of the second, is proportional
to the weak lensing angular correlation function. Although
at this stage it is difficult to give definitive quantitative
predictions, the magnitude of the fourth order correlation function should be
about $10^{-2}$ (the order of the corrective term in [\ref{corr2}]),
which should be easily detectable in full sky coverage
CMB maps. 
The expression of the four-point correlation function can be
given in terms of the power spectra,
\ba
&\mg\dTo(\gam_1)\,\dTo(\gam_2)\,\dTo(\gam_3)\,\dTo(\gam_4)\md_c=\\
&\disp{\int{\d^2\vl_1\over (2\pi)^2\,C_{l_1}}\,
\int{\d^2\vl_2\over (2\pi)^2\,C_{l_2}}}\,
\disp{\int\d\chi\,w^2(\chi)\,\int{\d^2\vk\over (2\pi)^2}\,P(k)}\nonumber\\
&\disp{{\vl_1\cdot\vk\over k^2\,\mD_0}\,{\vl_1\cdot\vk\over k^2\,\mD_0}\,
\exp\left[\ii\vl_1\cdot\gam_{12}+\ii\mD_0\vk\cdot\gam_{23}+
\ii\vl_2\cdot\gam_{34}\right]}+\nonumber\\
&{\rm perm.}\ (11\ {\rm other\ terms}).\nonumber
\ea
This expression can be calculated with an integration
over the angles between $\vl_1$ and $\gam_{12}$
and $\vl_2$ and $\gam_{34}$ respectively. It yields the Bessel 
functions $J_1(l_1\,\gam_{12})$ and $J_1(l_2\,\gam_{34})$.
The results can thus be expressed in terms of the 
angular derivative of the two-point correlation
function,
\be
\disp{\d\over\d\gam}C(\gam)=-
\disp\int{l^2\d l\over 2\pi}\,C_l\,J_1(l\gam),
\ee
and with quantities associated with the
angular correlation of the displacement field,
\be
D_p(\gam)=
\disp{
\int\d\chi\,w^2(\chi)\,\int{k\,\d k\over 2\pi}\ {P(k)\over k^2\,\mD_0^2}}\ 
\disp{J_p(\mD_0\,k\,\gam)},\label{Dp}
\ee
leading to (Appendix A),
\ba
&\mg\dTo(\gam_1)\,\dTo(\gam_2)\,\dTo(\gam_3)\,\dTo(\gam_4)\md_c=\\
&\disp{{1\over 2}
{\d\over\d\gam}C(\gam_{12})\,\disp{\d\over\d\gam}C(\gam_{34})\,
\times}\nonumber\\
&\disp{\left[D_0(\gam_{23})\,\cos(\varphi_{12}-\varphi_{34})-
D_2(\gam_{23})\,\cos(\varphi_{12}+\varphi_{34})\right]}+\nonumber\\
&+\ {\rm perm.}\ (11\ {\rm other\ terms}),\label{corr4}\nonumber
\ea
where $\varphi_{12}$ is the angle between $\gam_{12}$ and $\gam_{23}$\
and $\varphi_{34}$ is the angle between $\gam_{43}$ and $\gam_{32}$
(see Fig. 1).
Two terms are thus involved. The a priori dominant term is the one in
$D_0$, and it is weighted by the 
cosine of the angle $\varphi_{12}-\varphi_{34}\equiv\psi$ 
(see Fig. 1), that is the angle between 
the directions $\gam_{12}$ and $\gam_{34}$ on the sky. It gives a clear
geometrical dependence for the four point-correlation function. However,
one should have in mind that 11 other terms have to be taken 
into account in this calculation. This signal may therefore be
masked by other geometric dependences.

\begin{figure}
\vspace{3 cm}
\special{hscale=60 vscale=60 voffset=0 hoffset=10 psfile=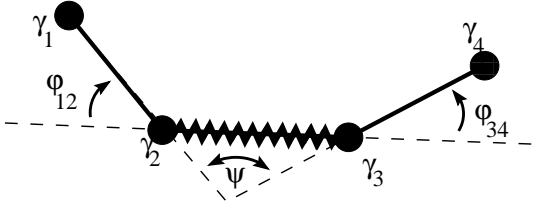}
\caption{Description of the angles intervening in the
expression (\ref{corr4}) of the four-point correlation function. The thick 
solid line materializes the $\d\,C(\theta)/\d\,\theta$ factor, whereas the 
hatched lines represents the correlation functions
of the displacement field.}
\end{figure}

Quantitative calculations can be done for specific cosmological
models (see next section). However, one obvious problem
for a practical determination of this correlation function is that
it depends on 5 different variables. It is thus crucial to reduce
the number of variables by considering simplified geometries.

\subsection{Peculiar Geometries}

\subsubsection{When two Directions Coincide}

The first geometry one may think of is when two points are merged together,
that is the expression of
$\mg\dTo(\gam_1)\,\left[\dTo(\gam_2)\right]^2\,\dTo(\gam_3)\md_c$. 
This notation is actually a bit oversimplified since the local
temperature fluctuations are actually filtered by the 
used apparatus. One should thus have in mind that the two directions
denoted $\gam_2$ are actually close random directions in a beam 
centered on $\gam_2$.

Of course, once again, many terms are contributing to this ensemble average 
but I will first concentrate on the case where the connection between 
the two $\dTo(\gam_2)$ is made by the lens coupling term (subsection A.2.b). 
In such a case one can see that $\psi$ is given by the
angle between $\gam_2-\gam_1$ and $\gam_3-\gam_2$
and is not affected by the smoothing. This is not the
case for the term in $\cos(\varphi_{12}+\varphi_{34})$
which is expected to vanish because it is averaged to zero
(more precise derivations are given in the Appendix).
This contribution is thus proportional to the cosine of the angle,  
and to the autocorrelation function of the displacement field.

\begin{figure}
\vspace{7 cm}
\special{hscale=45 vscale=45 voffset=0 hoffset=10 psfile=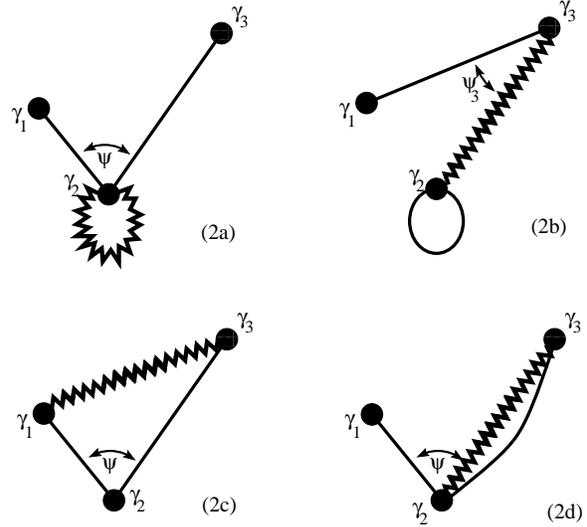}
\caption{Geometrical representation of the terms intervening
in the expression (\ref{G1K}).}
\end{figure}

What about the other terms? Their geometrical representations 
are given in Fig. 2. The (2a) diagram is the term that has just been 
considered and is expected to dominate the final expression. 
Note that all these diagrams have a symmetry factor of 2 compared to what is
given in the Appendix.
At first view the (2b) diagram vanishes because $\varphi_{12}$
takes a random value averaging both $\cos(\varphi_{12}-\varphi_{34})$
and $\cos(\varphi_{12}+\varphi_{34})$ to zero. A more detailed
calculation is proposed in the Appendix. It shows that
it gives a contribution proportional to the angular correlation function 
of the local weak lensing convergence (see 
Blandford et al. 1991, Villumsen 1996, Bernardeau et al. 1996),
\be
M_0(\gam)=\disp{\int\d\chi\,w^2(\chi)\,
\int{k\,\d k\over 2\pi}\,P(k)\,J_0(\mD_0\,k\,\gam)}.
\ee
The two other diagrams are simpler since they 
can be simply obtained from the general expression (\ref{corr4}).
Taking all these terms into account we have,
\ba
&\mg\dTo(\gam_1)\,\left[\dTo(\gam_2)\right]^2\,\dTo(\gam_3)\md_c
\approx\label{G1K}\\
&\disp{{\d\log(\Cb)\over \d\log(\theta_0)}\,{\d\over\d\gam}\,C(\gam_{13})\ 
\left[\gam_{12}\,M_0(\gam_{12})\,\cos(\psi_1)+\right.}\nonumber\\
&\disp{\left.\gam_{23}\,M_0(\gam_{23})\,\cos(\psi_3)\right]}+\nonumber\\
&\disp{\d\over\d\gam}C(\gam_{12})\,\disp{\d\over\d\gam}C(\gam_{23})\,
\cos(\psi)\times\nonumber\\
&\left[
\overline{D}_0(\theta_0)+D_0(\gam_{13})-D_0(\gam_{12})
-D_0(\gam_{23})\right]\nonumber
\ea
where $\psi_3$ is shown on Fig. (2b) ($\psi_1$ corresponds
to the diagram obtained when the roles of $\gam_1$ and $\gam_3$ are
inverted) and $\theta_0$ is the smoothing angle of the experiment.
In this expression, the $D_2$ terms have
been neglected, $D_0$ has been changed in $\overline{D}_0$ for 
diagram (2a) and $C$ in $\Cb$ in (2b). 
This is due to the filtering effects. One should indeed
take into account the average displacement within 
the beam size of the experiment in the first case, the rms 
temperature fluctuations in the second. More precisely we have,
\be
\overline{D}_0(\theta)=
\disp{
\int\d\chi\,w^2(\chi)\,\int{k\,\d k\over 2\pi}\ {P(k)\over k^2\,\mD_0^2}}\ 
\disp{W^2(\mD_0\,k\,\theta)},\label{Db0}
\ee
and
\be
\Cb(\theta)=\disp{\int{l\,\d l\over 2\pi}\,C_l\,W^2(l\,\theta)},
\ee
with, for a top-hat window function for instance,
\be
W(x)=\disp{2 J_1(x)\over x}.
\ee

It is then interesting to define the function, 
\ba
&\kappa_4(\gam_{12},\gam_{23})=\label{k4G2}\\
&\disp{
\mg\dTo(\gam_1)\,\left[\dTo(\gam_2)\right]^2\,\dTo(\gam_3)\md_c
\over
\mg\dTo(\gam_1)\,\dTo(\gam_2)\md\,\mg\dTo(\gam_2)\,\dTo(\gam_3)\md},\nonumber
\ea
which is a dimensionless quantity. It does not depend
in particular on the magnitude of the CMB temperature fluctuations
and it is directly proportional to the large-scale structure
power spectrum, with a known dependence on the shape of the
anisotropy power spectrum. 
This quantity would thus be a measure the weak lensing effects in CMB maps.
Taking into account the fact that the diagram (2b)
generally leads to a negligible contribution we have,
\ba
&\kappa_4(\gam_{12},\gam_{23})\approx\label{k4G2e}\\
&\disp{\d\over\d\gam}\log[C(\gam_{12})]\,
\disp{\d\over\d\gam}\log[C(\gam_{23})]\,
\cos(\psi)\times\nonumber\\
&\left[
\overline{D}_0(\theta_0)+D_0(\gam_{13})-D_0(\gam_{12})
-D_0(\gam_{23})\right].\nonumber
\ea
A 2D contour plot of the function $\kappa_4(\gam_{12},\gam_{23})$
is proposed in Fig. 6 for a peculiar cosmological case.

The main contribution to this expression is the term
coming from the diagram (2a), but the contributions from (2c) and (2d)
cannot be neglected because the correlation function of the displacement
field $D_0$ is only slowly decreasing with the angle
(see next section). 
It is interesting to have in mind the physical effect 
described by this result. It corresponds indeed to a shear effect
that is the {\em deformation of two nearby temperature patches by a single
lens}. It can thus be seen as the excess of 
close temperature peaks that are elongated in the {\sl same} direction.
It is therefore logical that this effect is proportional to the
correlation function of the displacement field. 
However, it does not always dominated the four point
correlation function. This is the case in particular at the
degree scale since the derivative of the temperature correlation
function vanishes (see Fig. 4 in the next Section).
The next subsection is devoted to a simpler geometry where
this case is more specifically  investigated.

\subsubsection{When two Pairs Coincide}

\begin{figure}
\vspace{6 cm}
\special{hscale=50 vscale=50 voffset=0 hoffset=10 psfile=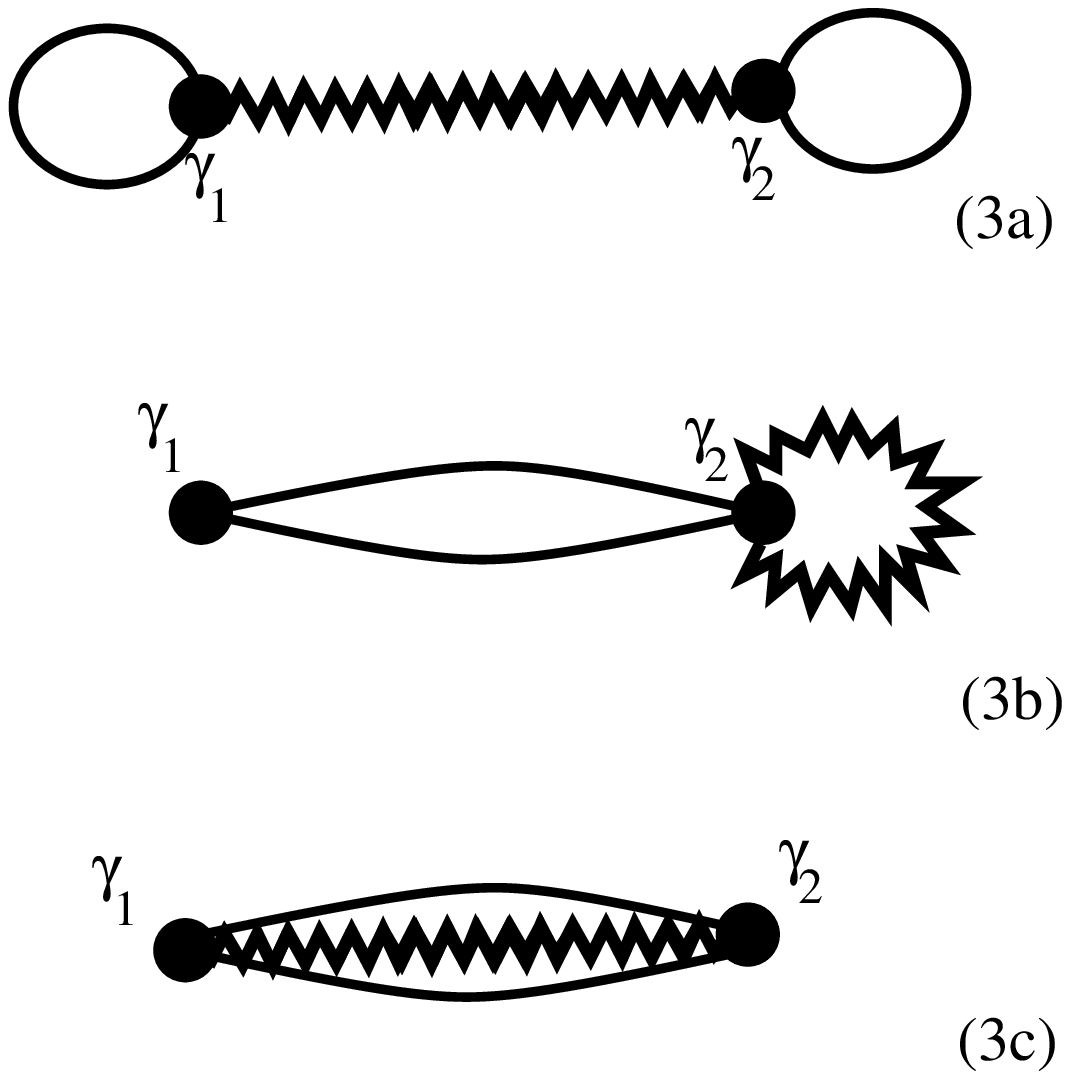}
\caption{Geometrical representation of the terms intervening
in the expression (\ref{G2K}).}
\end{figure}

This case is obtained from the  previous case when the directions $\gam_3$
and $\gam_4$ coincide. The geometrical representations of the
involved terms are presented in Fig. 3. The a priori dominant term
corresponds to the diagram (3b). However, as noted previously,
both the diagrams (3b) and (3c) 
vanish for specific values of the distance between $\gam_1$ and $\gam_2$. 
In such a case we are left with the diagram (3a). A crude
evaluation of this diagram gives zero because the averages
over the angles $\varphi_{12}$ and $\varphi_{34}$ are
expected to vanish. However this is true only when the smoothing
angle is negligible compared to the
angular distance between $\gam_1$ and $\gam_2$.
More precise calculations (subsection A.3) show that this term
is proportional to the logarithmic derivative of the temperature
variance in the beam. The resulting value of the correlation function
is (taking into account a symmetry factor of 4),
\ba
&\mg\left[\dTo(\gam_1)\right]^2\left[\dTo(\gam_2)\right]^2\md_c
\approx\label{G2K}\\
&\disp{{1\over 4}\left(\theta_0{\d\Cb(\theta_0)\over\d\theta}\right)^2}\ 
M_0(\gam_{12})+\disp{\left[{\d\over\d\gam}C(\gam_{12})\right]^2\times}
\nonumber\\
&\,\left[\overline{D}_0(\theta_0)-D_0(\gam_{12})\right].\nonumber
\ea
Note that the first term is no more proportional to the
angular correlation of the displacement field, but to the 
angular correlation of the convergence (that is, within a factor 2, 
of the magnification in the weak lensing regime). 
Here the physical mechanism has changed. The correlation
function is not due to a local shear, but to single lens amplification
of patches on the primordial sky that create an  
{\em excess of close bright peaks}. It is thus
clearly a magnification effect.

\subsubsection{When the four Directions Coincide}

The last case I consider is when the four directions coincide.
In this case the magnification effect always dominates and the results,
obtained in subsection A.4, are similar to the one discussed
in the previous subsection, but with small changes 
introduced by the filtering effects,
\be
\mg\left[\dTo(\theta_0)\right]^4\md_c={3\over4}
\disp{\left(\theta_0{\d\Cb(\theta_0)\over\d\theta}\right)^2}\ \M0b(\theta_0)
\ee
with
\be
\M0b(\theta_0)=\disp{\int\d\chi\,w^2(\chi)\,
\int{k\,\d k\over 2\pi}\,P(k)\,W^2(\mD_0\,k\,\theta_0)}.
\ee
This result has been properly demonstrated for a top-hat window function,
and should be roughly correct for other window functions.
Note that this expression is also the fourth cumulant
of the local filtered temperature probability distribution function.
We can as well define the dimensionless quantity,
\be
\kappa_4(\theta_0)=
\disp{\mg\left[\dTo(\theta_0)\right]^4\md_c\over
\mg\left[\dTo(\theta_0)\right]^2\md^2}\approx
{3\over4}\disp{\left({\d\log[\Cb(\theta_0)]\over\d\log[\theta]}
\right)^2}\M0b(\theta_0)\label{k4pdf}
\ee
which tells us that {\em the dimensionless kurtosis of the local filtered
CMB temperature probability distribution function is proportional
to the variance of the local filtered convergence.}

\section{Quantitative Predictions}

\subsection{The Cosmological Models}

The cosmological models used to illustrate the previous
results by quantitative predictions are
standard CDM models with $\Omega_{\rm baryon}=0.05$, $H_0=50\ km/s/$Mpc
with an initial Harrison-Zel'dovich spectrum.
Two cases have been chosen, $\Omega_0=1$, $\Lambda=0$ (model 1) and
$\Omega_0=0.3$, $\Lambda=0.7$ (model 2).
The transfer function and temperature power spectrum were
both computed with the code of Seljak \& Zaldarriaga (1996).

For convenience the mass fluctuation power spectra were 
approximated by simple analytic fits (similar to the ones
proposed by Bond \& Efstathiou 1984),
\be
P(k)=A\,\disp{k\over\left(1+\left[a\,k+(b\,k)^{3/2}+
(c\,k)^2\right]^u\right)^{2/u}},
\ee
with
\ba
u&=&1.13;\nonumber\\
a&=&\disp{6.5\over 3000\ \Gamma};\nonumber\\
b&=&\disp{3\over 3000\ \Gamma};\nonumber\\
c&=&\disp{1.7\over 3000\ \Gamma};\nonumber
\ea
and 
\ba
\Gamma&=&0.5\ \ {\rm for}\ \Omega_0=1.0\nonumber\\
\Gamma&=&0.13\ \ {\rm for}\ \Omega_0=0.3.\nonumber
\ea
The normalization factor is given by the relation (\ref{norm})
with observational constraints,
\be
l(l+1)\,C_l=\disp{24 \pi\over 5}\,\disp{\left({Q_0\over T_0}\right)^2}.
\ee
The measured values of $Q_0$ and $T_0$ by the
COBE satellite (Mather et al. 1994, Gorski et al. 1994)
\ba
T_0=&2.726\pm0.010\ K,\\
Q_0=&19.9\pm1.6\ 10^{-6}\ K,
\ea
provide the normalization constraint for $A$,
\be
A\approx1.01\ 10^{-8}\ \Omega_0^{-2}\ 
\disp{\left({a(\chicmb)\over D_+(\chicmb)}\right)^2}.\label{valA}
\ee
Note that in the previous equations, distance units have been chosen so that
$c\,H_0=1$.

\subsection{Numerical Results}

\begin{figure}
\vspace{5.5 cm}
\special{hscale=50 vscale=50 voffset=-90 hoffset=0 psfile=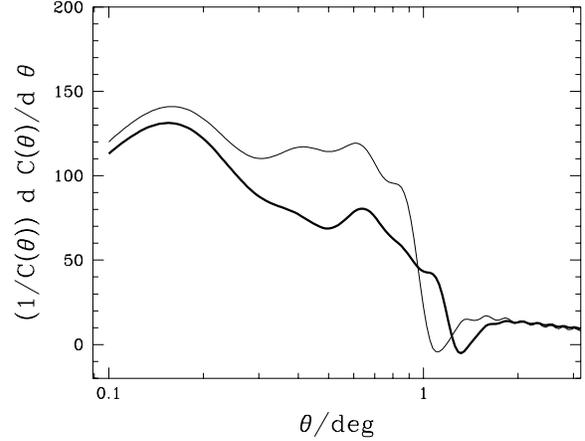}
\caption{The function $1/C(\theta)\ \d C(\theta)/\d\theta$
as a function of the angle. The thick line is for model
1 and the thin line for model 2}
\end{figure}

I present the derivatives of the temperature angular correlation
functions in Fig. 4. They exhibit a remarkable property, since they both 
drop to zero at a scale slightly larger than 1 degree. This property
can be of great help to disentangle various contributions.
For instance it implies that specific geometries produce a vanishing
four-point correlation function. This is the case for instance when
three of the four directions form an equilateral 
triangle with $\sim 1.2\,\deg$ of side length.
Such a property could be of crucial interest to ascertain the origin
of an observed four-point correlation function.

\begin{figure}
\vspace{5.5 cm}
\special{hscale=50 vscale=50 voffset=-90 hoffset=0 psfile=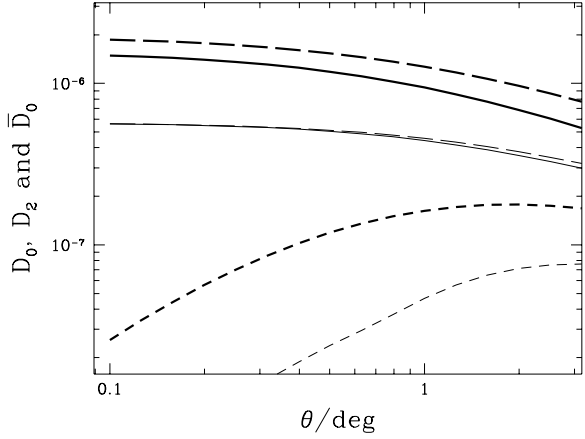}
\caption{The functions $D_0$ (solid lines), $D_2$ (dashed lines) 
and $\Db_0$ (long dashed lines) for model 1 (thick lines) and
model 2 (thin lines).}
\end{figure}

\begin{figure}
\vspace{6 cm}
\special{hscale=80 vscale=80 voffset=-30 hoffset=0 psfile=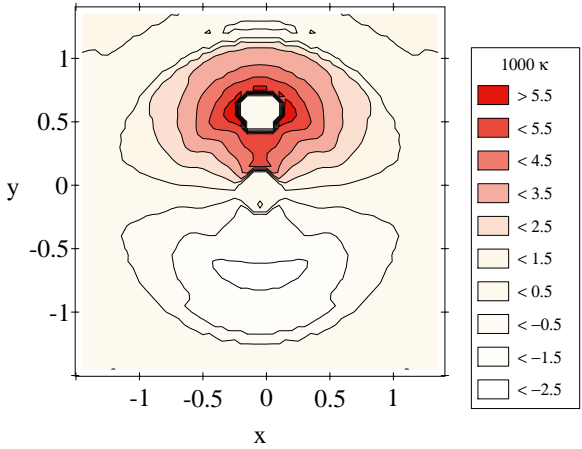}
\caption{Contour plot of the function $\kappa_4(\gam_{12},\gam_{23})$
(eq. \ref{k4G2}) as a function of the relative position (in degrees)
of $\gam_3$ when $\gam_2$ is the central point of the graph and
$\gam_1$ is at the coordinates $x=0$, $y=+0.63\deg$. The value of $\kappa_4$
has been multiplied by 1000.}
\end{figure}

The two other quantities of interest are
the functions $D_0(\gam)$, $D_2(\gam)$ and $\Db_0(\gam)$ (eqs. \ref{Dp},
\ref{Db0}) that describe
the magnitude of the lens effects. They are presented in Fig. 5.
One can see that $D_0$ and $\Db_0$ dominate at small scale. 
The resulting shape of the dimensionless four-point correlation
function $\kappa_4(\gam_{12},\gam_{23})$ (eq. \ref{k4G2}) is presented
in Fig. 6. It exhibits specific features induced by the
$\cos(\psi)$ factor and by the derivative of the 
temperature angular correlation function. 
The fact that the latter vanishes is clearly 
present with a significant circular feature
at 1 degree scale. It implies that the kurtosis is maximum
for angular distances below 
1 degree. Note that in this figure, all contributions have been
included, but the contributions from (2b) are found to be
negligible and not to affect the global features of this plot.

I also present the  quantities intervening in the expression of the
dimensionless kurtosis of the CMB temperature PDF. The index of the local
temperature fluctuations is given in 
Fig. 7 for the two cosmological models. The variance
of the convergence is presented in Fig. 8, and the resulting value
of $\kappa_4$ in Fig. 9.

\begin{figure}
\vspace{5.5 cm}
\special{hscale=50 vscale=50 voffset=-90 hoffset=0 psfile=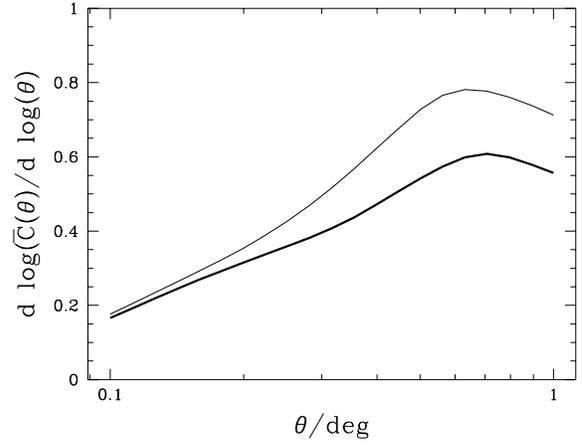}
\caption{The functions  $\d\log[\Cb(\theta)]/\d\log[\theta]$
for model 1 (thick line) and model
2 (thin line)}
\end{figure}

\begin{figure}
\vspace{5.5 cm}
\special{hscale=50 vscale=50 voffset=-90 hoffset=-10 psfile=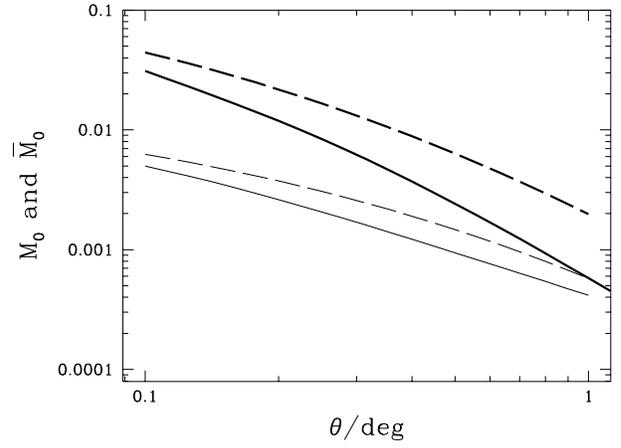}
\caption{The functions $M_0$ (solid lines) and
$\M0b$ (dashed lines) for model 1 (thick lines) and model
2 (thin lines)}
\end{figure}

\begin{figure}
\vspace{5.5 cm}
\special{hscale=50 vscale=50 voffset=-90 hoffset=0 psfile=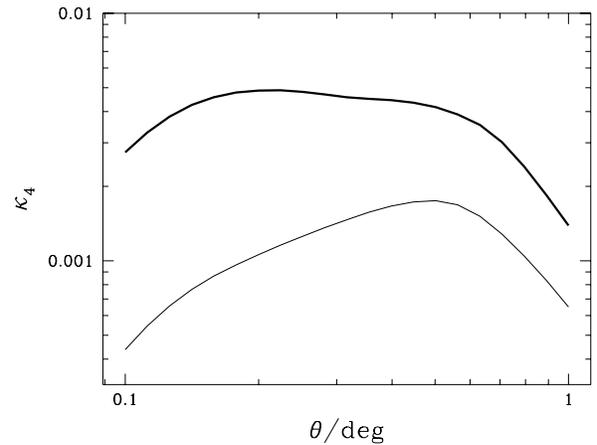}
\caption{The function $\kappa_4(\theta)$
(eq. \ref{k4pdf}) for model 1 (thick line) and model
2 (thin line)}
\end{figure}

The resulting coefficient $\kappa_4$ can be as high as $5\ 10^{-3}$.
It depends however a lot on the cosmological models and more particularly
on the amount of power at about $10\,h^{-1}$Mpc scale. The reason why $\M0b$,
and consequently $\kappa_4$, is smaller for the $\Omega=0.3$ case 
is thus actually due to the change of shape of $P(k)$
(i.e., a lower value of $\Gamma$), and not directly to the low value 
of $\Omega$. The dependence on the cosmological parameters is 
investigated in more details in the next subsection.

\subsection{Dependence on the Cosmological Parameters}

The dependence on the cosmological parameters enters in various
aspects. It is important in particular for the relation between the
$C_l$ and the mass fluctuation power spectrum $P(k)$.
This is not the aim of this paper to explore in great details all these
dependences. There is however a dependence which is specific of the
lens effects, that is the shape of the efficiency function
$w(z)$ (eq. \ref{eff}). In Fig. 10, I present
the function $w(z)$ for different cosmological models.
We can see that both the shape and the amplitude of this 
function strongly depend on the cosmological parameters.

The main dependence is actually due to the factor $\Omega$ 
appearing in the expression of $w(z)$. This is however quite 
misleading since this factor cancels out with the one intervening
in the expression of the normalization factor $A$ (eq. \ref{valA}).
The dependence on the cosmological parameters is thus mainly 
contained in the shape of the power spectrum, that is in the ratio
between the amount of power at relatively small scale to the power
at very large scale, although they enter significantly in the
efficiency function $w(z)$, even when the overall $\Omega$ factor
has been dropped. The way the two cosmological parameters
$\Omega_0$ and $\Lambda$ enter in this expression is unfortunately rather
cumbersome. It is however clear that the CMB four-point
correlation function contains informations on the cosmological 
parameters in a quite different combination compared to the 
one intervening in the temperature power spectrum. In particular
the detection of such an affect may allow to better disentangle
what is specific of $C_l$ from what contribute directly to $P(k)$.

\begin{figure}
\vspace{5.5 cm}
\special{hscale=50 vscale=50 voffset=-90 hoffset=0 psfile=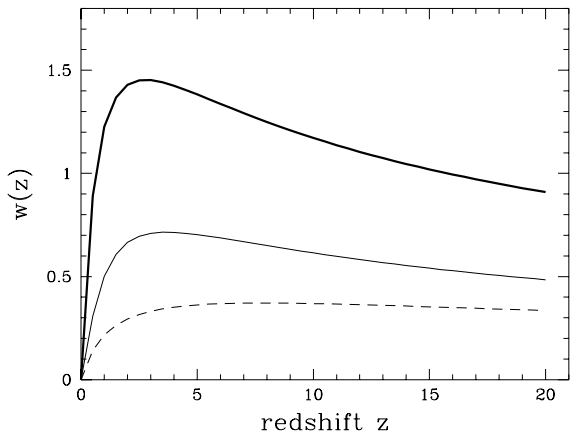}
\caption{The function $w(z)$ (eq. \ref{eff}) for the Einstein-de Sitter
case (thick solid line), a model with $\Omega=0.3$, $\Lambda=0.7$
(thin solid line) and a model with $\Omega=0.3$, $\Lambda=0$ (dashed line).}
\end{figure}

\section{Discussion and Conclusions}

In this work,
I have calculated the expression of the four-point temperature
correlation function as induced by weak-lensing effects.
For standard CDM model, the amplitude of this correlation
function, in units of the square of the second,
is found to be of order of $5\ 10^{-3}$. However, this estimation
did not take into account the nonlinear evolution of the power 
spectrum that might significantly amplify this signal at small 
angular scale.
This is for instance what is predicted for the two-point correlation
function of the polarization of background galaxies (Jain \& Seljak 1996).
Unfortunately, in the case of CMB maps, the scale at which this effect
might appear cannot be deduced straightforwardly from
this work.
This effect is indeed the result of a line of sight 
integration that mix different scales and different redshifts
for a given selection function (e.g. Fig. 10) which is itself dependent on
the redshift of the sources. Moreover all the
intervening quantities have non trivial dependences
on the cosmological parameters that should be taken into 
account. A detailed examination of the nonlinear effects is then
left for a forthcoming paper.

I would like to stress, that the amplitude of the lensing effects
should be large enough to be detectable, at least marginally,
in full sky CMB anisotropy measurements.
The possibility of doing such measurements is directly related
to the cosmic noise associated with the quantities of interest,
fourth moment or four-point correlation function. 
So far, no precise estimation of the cosmic noise for the four-point 
correlation function has been made, but following Srednicki (1993),
who presented the calculation for the three-point correlation
function, one expects the cosmic variance of those quantities
to be of the order of $1/l_{\rm pk.}$ where $l_{\rm pk.}$
is the typical value of $l$ contributing to the temperature
fluctuations. One can observe that the value of $1/l_{\rm pk.}$
is of the order of the signal, however, one should have in mind that
a direct and too naive calculation of the cosmic noise
may be actually misleading
since the  long wavelength fluctuations (corresponding to the low
$l$ part of the power spectrum) contribute significantly
to it, whereas the lensing signal 
originates mainly from the small angular scales (below 1 degree). 
It suggests that the weak lensing 
signal might be more easily detectable 
in maps where the long wavelength temperature
fluctuations have been removed. Moreover the detailed methods
used to extract the signal might also be of different robustness
against the cosmic noise. In particular, it could be
fruitful to take advantage of the a priori knowledge of the
geometrical dependence of the four-point correlation function (see
for example the $cos(\psi)$ factor in the expression [\ref{k4G2e}]).
In a forthcoming paper, we explore the different possible strategies
for the data analysis, and will
present detailed estimations of the precision at which such a detection
could be made in the future satellite missions.

It also has to be noted that other secondary effects or
foregrounds may also contribute to the 
four-point correlation function, not to
mention the case of more exotic cosmological models
based on intrinsically non-Gaussian topological defects. 
In particular the nonlinear Doppler effects could
induce a significant four-point correlation function, 
because it is caused by intrinsically non-Gaussian objects. There are however
no reasons for these effects to have the same geometrical
dependences, $\cos(\psi)$ factor and dependence 
on the shape of the temperature two-point correlation function.
Hence, it should be possible to distinguish this
effect from other sources.

The most exciting aspect of this analysis is
probably that the magnitude of the effect
depends on the cosmological parameters, $\Omega$,
$\Lambda$ and $P(k)$ in a known way. The detection of the
temperature four-point correlation function may thus reveal to be extremely
precious to test the global picture of the large-scale 
structure formation, as it will be unveiled
by CMB anisotropy measurements.

\section*{Acknowledgments}

The author is grateful to Fran\c cois Bouchet and Yannick Mellier 
for encouraging discussions, and to Uro\v s Seljak for the use
of his codes.

%\onecolumn
\section*{Appendix A: Calculation of the Filtering Effects} 

In this appendix I explicitly take into account the
filtering effects to compute the expressions of the 
four-point correlation function in different geometries. 
The filtering can be the one due to the angular resolution of the
apparatus or due to a subsequent filtering of the temperature field.
When it needs to be specified the adopted window function will
always be the angular top-hat window function.

The expression I am interested in is the expression (\ref{expc4}) where 
$\gamt$ and the coupling term $\nabla\gamt\cdot\dgam$
are given by,
\be
\dTt(\gam)=\disp{
\int{\d^2\vl\over (2\pi)^2}\,a_l\,W(l\,\theta_0)\,e^{\ii\vl\cdot\gam}},
\ee
and,
\ba
&\nabla\gamt(\gam)\cdot\dgam=\\
&\disp{\int{\d^2\vl\over (2\pi)^2}\,a_l\,W(l\theta_0)\,
\int\d\chi\,w(\chi)\,\int{\d^2\vk\over(2\pi)^2}\times}\nonumber\\
&\disp{
a_l\,\delta(\vk)\,{\vl\cdot\vk\over k^2\,\mD_0}\,
W\left(\vert\vl+\mD_0\,\vk\vert\,\theta_0\right)\,\
\exp[{\ii(\vl+\mD_0\vk)\cdot\gam}}],
\nonumber
\ea
where $\theta_0$ is the smoothing angle.
The quantity of interest is thus
\ba
&C_4(\gam_1,\gam_2,\gam_3,\gam_4)=
\disp{\int{\d^2\vl_1\over (2\pi)^2}\,C_{l_1}}\,
\disp{\int{\d^2\vl_2\over (2\pi)^2}\,C_{l_2}}\times\label{C4}\\
&\disp{\int\d\chi\, w^2(\chi)\,\int{\d^2\vk\over(2\pi)^2}
\,P(k)\,W(l_1\,\theta_0)}\times\nonumber\\
&\disp{{\vl_1\cdot\vk\over k^2\,\mD_0}\,
\exp[{\ii\vl_1\cdot\gam_{12}}]\,
W\left(\vert\vl_1-\mD_0\vk\vert\,\theta_0\right)\,
\exp[{\ii\mD_0\vk\cdot\gam_{23}}}]\nonumber\\
&\disp{W\left(\vert\vl_2-\mD_0\vk\vert\,\theta_0\right)\,
\exp[{\ii\vl_2\cdot\gam_{34}}]\,
{\vl_2\cdot\vk\over k^2\,\mD_0}\,W(l_2\,\theta_0).}\nonumber
\ea
In the following I will estimate this expression for different hypothesis
on $\gam_1$, $\gam_2$, $\gam_3$ and $\gam_4$.

\subsection*{A.1 Four separate Directions}

Here I assume that 
\be
\theta_0\ll\gam_{ij}\ \ {\rm whatever}\ i\ {\rm and}\ j.
\ee
The integral (\ref{C4}) will be dominated by values of $l_1$, $l_2$
and $k$ for which $l_1\,\gam_{12}$, $l_2\,\gam_{34}$
and $k\,\gam_{23}$ are about unity. It implies 
that $l_i\,\theta_0$ and 
$\vert\vl_i-\mD_0\vk\vert\,\theta_0$ are all small quantities
thus making the filtering effects negligible, so that,
\ba
&C_4(\gam_1,\gam_2,\gam_3,\gam_4)\approx
\disp{\int{\d^2\vl_1\over (2\pi)^2}\,C_{l_1}}\,
\disp{\int{\d^2\vl_2\over (2\pi)^2}\,C_{l_2}}\times\\
&\disp{\int\d\chi\, w^2(\chi)\,\int{\d^2\vk\over(2\pi)^2}
\,P(k)}\times\nonumber\\
&\disp{{\vl_1\cdot\vk\over k^2\,\mD_0}\,
\exp[\ii\vl_1\cdot\gam_{12}+\ii\mD_0\vk\cdot\gam_{23}+
\ii\vl_2\cdot\gam_{34}]\,
{\vl_2\cdot\vk\over k^2\,\mD_0}}.\nonumber
\ea
An interesting property to be used is that
\be
\disp{
\int{\d^2\vl\over (2\pi)^2}\,C_l\,\exp[\ii\vl\cdot\gam]\,
{\vl\cdot\vx\over l\,x}=-{\gam\cdot\vx\over \gam\,x}\,
\int{l\d l\over 2\pi}\,C_l\,J_1(l\,\gam)}.\label{PrCos}
\ee
It implies that
\ba
&C_4(\gam_1,\gam_2,\gam_3,\gam_4)=\\
&\disp{{\d C(\gam)\over\d\gam}(\gam_{12})\,
{\d C(\gam)\over\d\gam}(\gam_{34})}\,
\disp{\int\d\chi\, w^2(\chi)}\times\nonumber\\
&\disp{\int{\d^2\vk\over(2\pi)^2}
\,{P(k)\over k^2\,\mD_0^2}}\,\disp{{\gam_{12}\cdot\vk\over \gam_{12}\,k}\,
{\gam_{34}\cdot\vk\over \gam_{34}\,k}\,
\exp[\ii\mD_0\vk\cdot\gam_{23}]},\nonumber
\ea
with,
\be
C(\gam)=\disp{\int{\d^2\vl\over (2\pi)^2}\,C_l\,e^{\ii\vl\cdot\gam}}.
\ee
This integral can be eventually integrated
from properties of Bessel functions,
\ba
&\disp{\int_0^{2\pi}\d\phi\,\exp[\ii x \cos(\phi)]\,\cos^2(\phi)=\pi\,
\left[J_0(x)-J_2(x)\right],}\\
&\disp{\int_0^{2\pi}\d\phi\,\exp[\ii x \cos(\phi)]\,\sin^2(\phi)=\pi\,
\left[J_0(x)+J_2(x)\right],}\\
&\disp{\int_0^{2\pi}\d\phi\,\exp[\ii x \cos(\phi)]\,\sin(\phi)\,\cos(\phi)}
=0,
\ea
with which we find,
\ba
&\mg\dTo(\gam_1)\,\dTo(\gam_2)\,\dTo(\gam_3)\,\dTo(\gam_4)\md_c=\\
&\disp{{1\over 2}
{\d\over\d\gam}C(\gam_{12})\,\disp{\d\over\d\gam}C(\gam_{34})\,
\times}\nonumber\\
&\disp{\left[D_0(\gam_{23})\,\cos(\varphi_{12}-\varphi_{34})-
D_2(\gam_{23})\,\cos(\varphi_{12}+\varphi_{34})\right]},\nonumber
\ea
where,
\be
D_p(\gam)=
\disp{
\int\d\chi\,w^2(\chi)\,\int{k\,\d k\over 2\pi}\ {P(k)\over k^2}}\ 
\disp{J_p(\mD_0\,k\,\gam)},
\ee
$\varphi_{12}$ is the angle between $\gam_{12}$ and $\gam_{23}$\
and $\varphi_{34}$ is the angle between $\gam_{43}$ and $\gam_{32}$
(see Fig. 1).

\subsection*{A.2 When two Directions Coincide}

\subsubsection*{A.2.a When  $\gam_1$=$\gam_2$}

In this case,  $l_1$ is expected to be of the order of $1/\theta_0$,
thus larger than $k$ so that
\ba
&W\left(\vert\vl_1-\mD_0\vk\vert\,\theta_0\right)\approx
\disp{
W(\vl_1\,\theta_0)-}\\
&\disp{{\vl_1\cdot\vk\over l_1}\,\theta_0\,\mD_0\,W'(\vl_1\,\theta_0)
+...}\nonumber
\ea
As a result one has,
\ba
&C_4(\gam_1,\gam_1,\gam_2,\gam_3)=
\disp{\int{\d^2\vl_1\over (2\pi)^2}\,C_{l_1}}\,
\disp{\int{\d^2\vl_2\over (2\pi)^2}\,C_{l_2}}\\
&\disp{\int\d\chi\,w^2(\chi)\,\int{\d^2\vk\over(2\pi)^2}
\,P(k)}\times\nonumber\\
&\disp{
\exp[{\ii\mD_0\vk\cdot\gam_{12}}+{\ii\vl_2\cdot\gam_{23}}]\,
{\vl_2\cdot\vk\over k^2\,\mD_0}}\times\nonumber\\
&\disp{
\left(W^2(l_1\,\theta_0)\,{\vl_1\cdot\vk\over k^2\,\mD_0}-\theta_0\,
W'(l_1\,\theta_0)\,W(l_1\,\theta_0)\,{(\vl_1\cdot\vk)^2\over l_1\,k^2}\right)}.
\nonumber
\ea
When one integrates over the angle of $\vl_1$ the first term
vanishes. The second term of the expansion is thus the dominant
contribution, which takes the form,
\ba
&C_4(\gam_1,\gam_1,\gam_2,\gam_3)=
\disp{-{1\over 4}\,\theta_0{\d\Cb(\theta_0)\over\d\theta}}\,
\disp{\int{\d^2\vl_2\over (2\pi)^2}\,C_{l_2}}\\
&\disp{\int\d\chi\, w^2(\chi)\,\int{\d^2\vk\over(2\pi)^2}
\,P(k)}\times\nonumber\\
&\disp{
\exp[{\ii\mD_0\vk\cdot\gam_{12}}+{\ii\vl_2\cdot\gam_{23}}]\,
{\vl_2\cdot\vk\over k^2}}.\nonumber
\ea
Using the property (\ref{PrCos}) we have
\ba
&C_4(\gam_1,\gam_1,\gam_2,\gam_3)=-
\disp{{1\over 4}\,\theta_0{\d\Cb(\theta_0)\over\d\theta}\,
{\d\over\d\gam}C(\gam_{23})}\times\\
&\disp{\gam_{12}\cdot\gam_{23}\over\gam_{23}}
\,\disp{\int\d\chi\, w^2(\chi)}\times\nonumber\\
&\disp{\int{k\,\d k\over2\pi}
\,P(k)\,{J_0(\mD_0\,k\,\gam_{12})+J_2(\mD_0\,k\,\gam_{12})\over 2}}.\nonumber
\ea
Interestingly $C_4$ is now proportional to the
angular correlation function of the local {\sl magnification}
and not of the local displacement.

\subsubsection*{A.2.b When  $\gam_2$=$\gam_3$}

In this case  $k$ is expected to be of the order of $1/\theta_0$,
thus larger than $l_1$ and $l_2$.
As a result one has,
\ba
&C_4(\gam_1,\gam_2,\gam_2,\gam_3)=
\disp{\int{\d^2\vl_1\over (2\pi)^2}\,C_{l_1}}\,
\disp{\int{\d^2\vl_2\over (2\pi)^2}\,C_{l_2}}\\
&\disp{\int\d\chi\, w^2(\chi)\,\int{\d^2\vk\over(2\pi)^2}
\,P(k)}\times\nonumber\\
&\disp{
\exp[{\ii\vl_1\cdot\gam_{12}}+{\ii\vl_2\cdot\gam_{23}}]\,
{\vl_2\cdot\vk\over k^2\,\mD_0}}\disp{
W^2(k\,\theta_0)\,{\vl_1\cdot\vk\over k^2\,\mD_0}},\nonumber
\ea
leading to
\ba
&C_4(\gam_1,\gam_2,\gam_2,\gam_3)=
\disp{{1\over 2}{\d\over\d\gam}}
C(\gam_{12})\,\disp{\d\over\d\gam}C(\gam_{23})\times\\
&\cos(\psi)\,\overline{D}_0(\theta_0),\nonumber
\ea
where 
\be
\overline{D}_0(\theta)=
\disp{
\int\d\chi\,w^2(\chi)\,\int{k\,\d k\over 2\pi}\ {P(k)\over k^2\,\mD_0^2}}\ 
\disp{W^2(\mD_0\,k\,\theta)}.
\ee
Here the effect is proportional to the mean displacement
in the beam size $\theta_0$.

\subsubsection*{A.2.c Other Cases}

The other cases do not give specific formulae and can be derived from
results of section A.1.

\subsection*{A.3 When two Pairs Coincide}

\subsubsection*{A.3.a When $\gam_1=\gam_2$ and $\gam_3=\gam_4$}

This case is similar to the A.2.b case where the
result is dominated by the second order term of an expansion
in $k\,\theta_0$, that here should be written for $\vl_1$ and $\vl_2$.
We then have
\ba
&C_4(\gam_1,\gam_1,\gam_2,\gam_2)=
\disp{\int{\d^2\vl_1\over (2\pi)^2}\,C_{l_1}}\,
\disp{\int{\d^2\vl_2\over (2\pi)^2}\,C_{l_2}}\nonumber\\
&\disp{\int\d\chi\, w^2(\chi)\,\int{\d^2\vk\over(2\pi)^2}
\,P(k)}\times\\
&\disp{
l_1\theta_0\,W(l_1\,\theta_0)\,
W'(l_1\,\theta_0)\,\left({\vl_1\cdot\vk\over k\,l_1}\right)^2\,
\exp[{\ii\mD_0\vk\cdot\gam_{12}}]}\times\nonumber\\
&\disp{\left({\vl_2\cdot\vk\over k\,l_2}\right)^2\,l_2\theta_0\,
W(l_2\,\theta_0)\,W'(l_2\,\theta_0)}.\nonumber
\ea
The integrations over the angles between $\vl_1$ and $\vk$
and between $\vl_2$ and $\vk$ give each a factor $1/2$
leading to
\ba
&C_4(\gam_1,\gam_1,\gam_2,\gam_2)=
\disp{{1\over 16}\,\left(\theta_0{\d\Cb(\theta_0)\over\d\theta}\right)^2}
\times\\
&\disp{\int\d\chi\, w^2(\chi)\,\int{\d^2\vk\over(2\pi)^2}
\,P(k)}\,\disp{\exp[{\ii\mD_0\vk\cdot\gam_{12}}]},\nonumber
\ea
which can be expressed in terms of the angular correlation function of the 
magnification.

\subsubsection*{A.3.b When $\gam_1=\gam_4$ and $\gam_2=\gam_3$}

This case is a particular case of subsection A.2.a.

\subsection*{A.4 When the four Directions Coincide}

In this case we have
\ba
&C_4=\disp{\int{\d^2\vl_1\over (2\pi)^2}\,C_{l_1}}\,
\disp{\int{\d^2\vl_2\over (2\pi)^2}\,C_{l_2}}\,\disp{
\int\d\chi\, w^2(\chi)}\nonumber\\
&\disp{\int{\d^2\vk\over(2\pi)^2}
\,P(k)}\,{\vl_2\cdot\vk\over k^2\,\mD_0}\,{\vl_1\cdot\vk\over k^2\,\mD_0}
\times\\
&W\left(\vert\vl_1-\mD_0\vk\vert\,\theta_0\right)\,
W\left(\vert\vl_2-\mD_0\vk\vert\,\theta_0\right)\,
W(l_1\,\theta_0)\,W(l_2\,\theta_0).\nonumber
\ea
This expression cannot be simplified furthermore 
if the window function is not specified. 
In this paragraph I assume that $W$ is the top-hat window function,
\be
W(x)={2\ J_1(x)\over x}.
\ee
To complete the calculation it is interesting to have in mind
the property (Bernardeau 1995),
\ba
&\disp{\int{\d^2\vl_1\over (2\pi)^2}\,{\d^2\vl_2\over (2\pi)^2}\,
W\vert\vl_1+\vl_2\vert\,\left(1+{\vl_1\cdot\vl_2\over l_1^2}\right)=}
\label{propW}\\
&\disp{\int{l_1\d l_1\over 2\pi}\,
W(l_1)\,\int{l_2\d l_2\over 2\pi}\,
\left[W(l_2)+{1\over2}\,l_2\,W'(l_2)\right]}.\nonumber
\ea
This property is rigorously exact. In principle it is not possible to 
separate the two terms. Both relations are however good approximation as it
will be shown in the following. It is thus reasonable to assume that
\ba
&\disp{\int{\d^2\vl_1\over (2\pi)^2}\,{\d^2\vl_2\over (2\pi)^2}\,
W\vert\vl_1+\vl_2\vert\,{\vl_1\cdot\vl_2\over l_1^2}\approx}
\label{papW}\\
&\disp{{1\over 2}\int{l_1\d^ l_1\over (2\pi)}\,
W(l_1)\,\int{l_2\d l_2\over (2\pi)}\,l_2\,W'(l_2)}.\nonumber
\ea
Then, using this property it is easy to show that
\be
\disp{C_4={1\over 16}\,\left({\d\log[\Cb(\theta_0)]\over 
\d\log[\theta]}\right)^2\,\M0b(\theta_0)}.
\ee

\subsection*{A.5 Validity of the top-hat window function property (\ref{papW})}

To examine the property (\ref{papW}), an 
interesting property of the Bessel function to use is that
(Gradshteyn \& Ryzhik, 1980, eq. [8.532.1]),
\ba
&\disp{
{2\,J_1\left(\vert\vk_1+\vk_2\vert\right)\over \vert\vk_1+\vk_2\vert}=
\sum_{p=0}^{\infty}
(1+p)\,{2\,J_1(k_1)\over k_1}\,{2\,J_1(k_2)\over k_2}}\times\\
&\disp{C_p^1\left(-{\vk_1\cdot\vk_2\over k_1 k_2}\right)},\nonumber
\ea
with
\be
\disp{
C_p^1\big[-\cos(\varphi)\big]=
(-1)^{p+1}\,{\sin(p+1)\varphi\over \sin(\varphi)}}.
\ee
We have thus
\be
\int_0^{2\pi}\d\varphi\,C_p^1(-\cos(\varphi))=0
\quad{\rm for}\quad p\quad{\rm odd},
\ee
and
\be
\int_0^{2\pi}\d\varphi\,C_p^1(-\cos(\varphi))=2\pi
\quad{\rm for}\quad p\quad{\rm even}.
\ee
As a result
\ba
&\disp{
C_4=2^6\,\int{\d l_1\over 2\pi}\,C_{l_1}\,\int{\d k\over 2\pi}\,P(k)\,
\int{\d l_2\over 2\pi}\,C_{l_2}}\\
&\disp{J_1(l_1\,\theta_0)\,J_1(l_2\,\theta_0)\sum_{p_1,p_2}
(1+2p_1)\,(1+2p_2)}\times\nonumber\\
&J_{1+2p_1}(l_1\,\theta_0)\,
J_{1+2p_1}(k\,\theta_0)\,J_{1+2p_2}(k\,\theta_0)\,
J_{1+2p_2}(l_2\,\theta_0)\nonumber
\ea
In the following I assume a power law behavior for both the mass power 
spectrum and $C_l$,
\ba
P(k)&\propto\,k^{n_w},\\
C_l&\propto\,l^{n_c}.
\ea
Then using the property,
\ba
&\disp{\int_0^{\infty}J_{\nu}(t)\,J_{\mu}(t)\,t^{-\lambda}\d t=}\\
&\disp{
{\Gamma(\lambda)\over 2^{\lambda}}\,
{\Gamma\left({\nu+\mu-\lambda+1\over2}\right)\over
\Gamma\left({\nu+\mu+\lambda+1\over2}\right)\,
\Gamma\left({-\nu+\mu+\lambda+1\over2}\right)\,
\Gamma\left({\nu-\mu+\lambda+1\over2}\right)}},\nonumber
\ea
one can easily compute the few terms of the previous series. It is
found to converge very rapidly.
The resulting ratio,
\ba
&r(n_c,n_w)=
\sum_{p_1,p_2}\,(1+2p_1)\,(1+2p_2)\times\label{rat}\\
&\disp{{S(p_1,1,n_c)\,S(p_1,p_2,n_w)\,S(1,p_2,n_c)
\over S(1,1,n_c)^2\,S(1,1,n_w)}},\nonumber
\ea
with
\ba
&\disp{S(p_1,p_2,n)=}
\disp{{\Gamma\left(1\!+\!p_1\!+\!p_2\!+\!n/2\right)\,\over
\Gamma\left(p_2\!-\!p_1\!+\!1\!-\!n/2\right)}}\times\\
&\disp{{1\over
\Gamma\left(p_2\!+\!p_1\!+\!2\!-\!n/2\right)\,
\Gamma\left(-p_2\!+\!p_1\!+\!1\!-\!n/2\right)}},\nonumber
\ea
is plotting in Fig. 11.
It shows that the error made by using the approximation is at most
of a few percent for the values of $n_c$ and $n_w$
of interest.
\begin{figure}
\vspace{6 cm}
\special{hscale=80 vscale=80 voffset=-30 hoffset=0 psfile=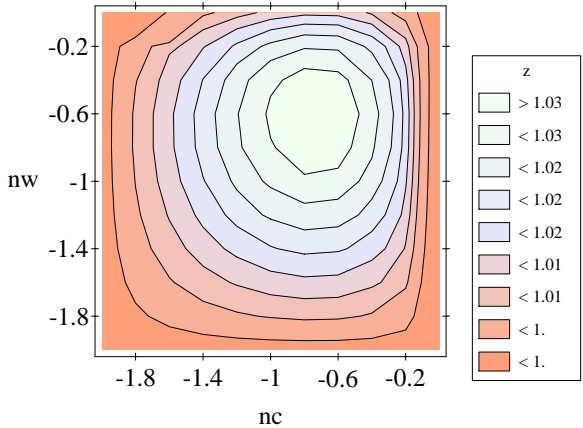}
\caption{The ratio (\ref{rat}) as a function of $n_c$ and $n_w$.}
\end{figure}

\end{document}